# Offline Algorithms for Several Network Design, Clustering and QoS Optimization Problems

Mugurel Ionut Andreica*, Eliana-Dina Tirsa*, Alexandru Costan*, Nicolae Tapus*

*Computer Science and Engineering Department, Politehnica University of Bucharest, Bucharest, Romania
(e-mail: {mugurel.andreica, eliana.tirsa, alexandru.costan, nicolae.tapus}@cs.pub.ro)

**Abstract:** In this paper we address several network design, clustering and Quality of Service (QoS) optimization problems and present novel, efficient, offline algorithms which compute optimal or near-optimal solutions. The QoS optimization problems consist of reliability improvement (by computing backup shortest paths) and network link upgrades (in order to reduce the latency on several paths). The network design problems consist of determining small diameter networks, as well as very well connected and regular network topologies. The network clustering problems consider only the restricted model of static and mobile path networks, for which we were able to develop optimal algorithms.

## 1. INTRODUCTION

Network design, clustering and Quality of Service (QoS) optimization problems arise in a wide range of fields (such as efficient data distribution and replication, providing QoS guarantees, and so on) and developing efficient algorithms for solving such problems is an important goal in computer science research. In this paper we address several such problems from an offline perspective. In Section 2 we discuss the issue of computing backup shortest paths in a network, when the last link on the shortest path may fail. In Section 3 we consider two problems regarding network link latency changes in order to satisfy several QoS constraints. In Section 4 we discuss 2 network design problems with the objectives of obtaining a bounded network diameter and a k-regular network topology. In Section 5 we consider several network clustering problems for the restricted case of path networks. In Section 6 we discuss related work and we conclude.

## 2. BACKUP SHORTEST PATHS

We consider a network composed of $n$ nodes and $m$ undirected edges (links). Each edge $(u,v)$ has a latency $l(u,v)$. One of the nodes ($src$) occasionally has to deliver pieces of content to the other nodes of the network. The content is delivered along the shortest path from node $src$ to the destination node $d$. We are interested in computing backup paths from the source node $src$ to every other node in the network, for the case that the last edge on the shortest path from $src$ to each node $d$ fails. The backup path from $src$ to a node $d$ is the shortest path between $src$ and $d$ in the graph obtained from the original network by deleting the last edge on the (initial) shortest path from $src$ to $d$. At first, we will compute the shortest path tree $SPT$ in $O(m \cdot log(n))$ time (or even $O(m+n \cdot log(n))$). This tree, which is rooted at the node $src$ contains all the shortest paths from $src$ to every other node. The unique path between $src$ and a node $d$ in $SPT$ is the shortest path between these two nodes in the original graph. Each node $d$ has a parent $parent(d)$ in $SPT$. The last edge on the shortest path between $src$ and a node $d$ is the edge $(parent(d),d)$. The level of a node $d$ is the number of edges on the path from the root $src$ to $d$: $level(src)=0$ and $level(d\neq src)=level(parent(d))+1$. The length of the shortest path from $src$ to a node $d$ is $SP(d)$. We will traverse the shortest path tree in a DFS manner, starting from the root and assign to each vertex $d$ its DFS number $DFSnum(d)$ ($DFSnum(d)=j$ if $d$ was the $j^{th}$ distinct vertex visited during the DFS traversal). We will then assign to each node $d$ an interval $I(d)=[DFSnum(d),DFSmax(d)]$, where $DFSmax(d)$ is the largest DFS number of a node in $T(d)$ (node $d$'s subtree). A simple method of computing backup paths is to recompute the shortest path in the graph between node $src$ and every node $d$, after removing the edge $(parent(d),d)$ from the graph (network). This approach takes $O(n \cdot m \cdot log(n))$ (or $O(n \cdot m+n^2 \cdot log(n))$) time, which may be too long when the number of connections is quite large (for instance, if $m=O(n^2)$, the time complexity can become as high as $O(n^3)$, which is prohibitive for networks composed of a large number of nodes). We will present two solutions, with $O(n^2+m)$ and $O((n+m) \cdot log(n))$ complexities. The first approach traverses the shortest path tree $SPT$ in a bottom-up manner. For each node $d$, it computes an array $BPL(d)$, where $BPL(d,j)$ ($0 \leq j < level(d)$) is the shortest length of a backup path which diverges from the shortest path from $src$ to $d$ at level $j$. We will initialize $BPL(d,j)$ to $+\infty$. Then, we consider every son $s(d,q)$ of the node $d$ ($1 \leq q \leq ns(d)$; $ns(d)$=the number of sons of node $d$ in $SPT$) and set $BPL(d,j)= min\{BPL(d,j), BPL(s(d,q),j)+l(d,s(d,q))\}$ ($0 \leq j < level(d)$). Afterwards, we consider all the edges $(u,d)$ and compute $LCA(u,d)$ (the lowest common ancestor of $u$ and $d$ in $SPT$). If $(u \neq parent(d))$ and ($d$ is not an ancestor of $u$ in $SPT$), then $BPL(d, level(LCA(u, d))) = min\{BPL(d, level(LCA(u, d))), SP(u)+l(u,d)\}$. Two methods for testing if $d$ is an ancestor of $u$ are to check if: (1) $(LCA(u,d)=d)$; or (2) $(DFSnum(u) \in I(d))$ (i.e. $DFSnum(d) \leq DFSnum(u) \leq DFSmax(d)$). There are many techniques for computing the lowest common ancestor, the fastest of which takes $O(n)$ preprocessing and $O(1)$ time per query (Bender et al., 2000). The length of the shortest backup path for every node $d$ is $BP(d)=min\{BPL(d,j)| 0 \leq j < level(d)\}$. In order to compute the actual backup paths we just need to trace back the way we computed the *BP(*)* and *BPL(*,*)* values. The second approach improves the first algorithm. During the bottom-up traversal, we will maintain a data structure *DS* which will be used as follows. When we consider all the edges *(u,d)* for a node *d*, if *(u≠parent(d))* and *(LCA(u,d)≠d)* then we add the tuple *(val=SP(u)+l(u,d)-SP(d), v=d, lca=LCA(u,d))* to *DS*. Afterwards, we compute *BP(d)= SP(d)+DS.Q(d)* (*DS.Q(d)*=the minimum value of all the tuples *(value, v, lca)*, where *lca* is an ancestor of *d* in *SPT* and *v* is a descendant of *d* in *SPT*). We have two choices for *DS*. *DS* can be a 2D dynamic range tree. When we add a tuple *(val, v, lca)* to *DS*, we insert a point *(DFSnum(v), level(lca))* with weight *val* in the range tree (in $O(log^2(n))$ time). *DS.Q(d)* computes in $O(log^2(n))$ time the minimum weight of a point in the 2D range *I(d)x[0, level(d)-1]* (or +∞ if no such point exists). The second choice is a segment tree *ST*. Every node *d* is associated to the leaf *DFSnum(d)* of *ST*. Adding a tuple *(val, v, lca)* to *DS* means adding it to a balanced tree *BT(r)* stored in the leaf *r=DFSnum(v)* from *ST*, as well as to a list *LT(lca)*. The current value assigned to a leaf of *ST* is the minimum value within its balanced tree (or +∞ if this tree is empty). The non-leaf nodes of *ST* store the minimum value assigned to any leaf in their subtree. As soon as the bottom-up traversal arrives at a node *d*, all the pairs *(val, v, lca=d)* from *LT(d)* are removed from *BT(r)* of the leaf *r=DFSnum(v)*; after every insertion/deletion of a tuple in/from some tree *BT(r)*, all the values assigned to the leaf *r* of *ST* and to its ancestors in *ST* are recomputed. Then, we call *DS.Q(d)* which computes the minimum value assigned to any leaf of *ST* in the interval *[DFSnum(d), DFSmax(d)]*. We perform *O(n)* queries and *O(m)* insertions/deletions into *BT(*)*, each of them taking *O(log(n))* time.

## 3. QoS-CONSTRAINED LINK LATENCY CHANGES

The first problem we consider is the following. We are given an undirected graph (network) with *n* nodes and *m* edges (links). Each edge *(u,v)* has a latency *l(u,v)≥0*. We want to solve an inverse optimization problem. We want to change the latency (increase it or decrease it) such that the shortest path from a source node *src* to every vertex *d* is exactly *SP(d)*. We want to minimize the sum of the values *|l'(u,v)-l(u,v)|* over all the edges *(u,v)*, where *l'(u,v)* is the new latency of the edge *(u,v)*. The new latency *l'(u,v)* must satisfy the constraint: *l'(u,v)≥lmin(u,v)* (initially, *l(u,v)≥lmin(u,v)*). At first, we will increase the latency of every edge *(u,v)* for which *SP(u)<SP(v)* and *SP(u)+l(u,v)<SP(v)*; the new latency of the edge *(u,v)*, *l'(u,v)*, will be *SP(v)-SP(u)*. For the other edges, we initialize *l'(u,v)* to *l(u,v)*. After this initial step, we will sort the vertices in increasing order of their distances from the source node *src* (we will also consider node *src*, with *SP(src)=0*). Each node *u* will have an associated cost *C(u)*. Initially, *C(src)=0* and *C(u≠src)=+∞*. We will insert all the tuples *(SP(i),C(i),i)* into a min-heap; we have *(SP(i),C(i), i)<(SP(j),C(j),j)* if *(SP(i)<SP(j))* or *((SP(i)=SP(j))* and *(C(i)< C(j)))*. As long as the heap contains any elements, we will extract the minimum element from the heap. We will also maintain an array *extracted*, which is initially set to *0* for all the nodes. Let's assume that we extracted the tuple *(SP(u),C(u),u)*. We will mark *u* as being extracted (we set *extracted(u)* to *1*). Afterwards, we will consider every edge *(u,v)*. If *(SP(u)≤SP(v))* and *(extracted(v)=0)* and *(lmin(u,v)≤ (SP(v)-SP(u)))* and *(l'(u,v)-(SP(v)-SP(u))<C(v))* then we remove the tuple *(SP(v),C(v),v)* from the min-heap, set *C(v)* to *(l'(u,v)-(SP(v)-SP(u)))* and insert the tuple *(SP(v),C(v),v)* (with the modified value *C(v)*) back into the min-heap; we also set *parent(v)=u*. When the heap becomes empty, we traverse all the vertices *u≠src* and set *l'(parent(u),u)* to *SP(u)- SP(parent(u))*. If any vertex *v* still has *C(v)=+∞*, then no solution exists. The time complexity is *O((n+m)·log(n))*. A more applicative version of the problem we have just described is the following. Let's assume that we are given the same graph as before, but we want that the latency of the shortest path from a source node *src* to every node *d* is at most *SP(d)*. In order to solve this problem we compute the shortest path from *src* to every other vertex *d* (let *SPlen(d)* denote the length of this shortest path). If *SP(d)>SPlen(d)*, we set *SP(d)=SPlen(d)*. Afterwards, we solve the problem described previously, with the new values *SP(*)*. This way, the initial step of the algorithm presented above, where the latency of some edges is increased, is not necessary anymore.

In the second problem we are given a (multicast) tree with *n* vertices, rooted at a source vertex *src*. The latency of each edge *(u,v)* is *l(u,v)≥0*. We want to decrease the latencies of the edges to some new values *l'(u,v)*, such that the maximum distance from *src* to every other vertex is as small as possible. Moreover, the cost, which is represented by the sum of the values *(l(u,v)-l'(u,v))* (over all the edges *(u,v)*) should be at most *C*. An extra condition is that the latency of an edge *(u,v)* can be decreased at most down to *lmin(u,v)≥0* (i.e. *lmin(u,v)≤l'(u,v)≤l(u,v)*). We will traverse the tree and assign to each vertex *i* its DFS number and then compute the interval *I(i)* (defined previously). We will compute the distance from *src* to every vertex of the tree: *d(src)=0* and *d(i≠src)=l(parent(i),i)+ d(parent(i))*. Let's consider the vertices *v(1), …, v(n)*, in increasing order of their DFS numbers. We will construct a segment tree *A* (Andreica et al., 2008) over all the *n* vertices, sorted according to their DFS numbers (the segment tree will have *n* leaves). The value assigned to every leaf *i* of the segment tree will be *d(v(i))*. The internal nodes of the segment tree will maintain the maximum value of a leaf in their subtrees. We will construct another segment tree *B* over the tree vertices (considered in the same order), where we will perform range set updates. Initially, for every son *s(src,j)* (*1≤j≤ns(src)*) of the root node *src*, we will range update the interval *I(s(src,j))* in *B*, by setting all the values in the corresponding interval to *s(src,j)*. We will also maintain a counter $C_{total}$=the total cost spent during the algorithm (initially, $C_{total}=0$). We initialize the latencies *l'(u,v)* to *l(u,v)*. A first approach (which works for integer latency values) proceeds as follows. As long as $C_{total}<C$, we perform the following actions: we query the segment tree *A* and find the leaf *i* with the largest value assigned to it. Then, we query the segment tree *B*, in order to find the vertex *x* to which the leaf *i* was set by the most recent range set update. The edge *(x,parent(x))* is the edge whose latency will be decreased by *1* unit (we assume that the latencies are integers), if possible. If *l'(x,parent(x))> lmin(x,parent(x))*, we set *l'(x,parent(x))= l'(x,parent(x))-1*; afterwards, we range update the interval *I(x)* in the segment tree *A*, by decreasing by *1* the values assigned to the leaves in the interval *I(x)*; we also increase $C_{total}$ by *1*. If, instead, *l'(parent(x),x)=lmin(parent(x),x)* and *x* is not a leaf in the tree, we will consider every son *s(x,j)* ($1 \leq j \leq ns(x)$) of the vertex *x* and range set all the values in the interval *I(s(x,j))* of the segment tree *B* to *s(x,j)*; if *l'(parent(x),x)= lmin(parent(x),x)* and *x* is a leaf in the tree, then the algorithm stops and the maximum distance is the one corresponding to the vertex *v(i)*. This algorithm has time complexity $O((C+n) \cdot log(n))$, because every (range) query and every (range) update can be performed in *O(log(n))* time. From an implementation point of view, we will use the segment tree algorithmic framework introduced in (Andreica et al., 2008). We could also use a block partition instead of a segment tree, but the time complexity would drop to $O((C+n) \cdot sqrt(n))$. In the second approach we will binary search the minimum maximum distance from the source node *src* to every other vertex of the tree in the interval *[0,DMAX=max{d(i)|1≤i≤n}]*. In order to perform the feasibility test for a candidate distance *D*, we will traverse the tree vertices *i* in any order; for each vertex *i*, we will use the segment tree *A* in order to compute the current distance *CD* from the root to the vertex *i* (by point-querying the value assigned to the leaf *DFSnum(i)* in the segment tree *A*). While *(CD>D)* we perform the following actions. Just like in the previous algorithm, we query the segment tree *B* in order to find the last value *x* to which the leaf *DFSnum(i)* was set. Then, if *(l'(parent(x),x)-(CD-D)≥lmin(parent(x),x))*, we decrease *l'(parent(x),x)* by *(CD-D)*, we increase $C_{total}$ by *(CD-D)* and we range decrease the values in the interval *I(x)* of the segment tree *A* by *(CD-D)*; afterwards, we query the distance *CD* again for the leaf *DFSnum(i)* from the segment tree *A* – it should be equal to *D*. If, instead, *(l'(parent(x))-(CD-D)<lmin(parent(x),x))*, we decrease *l'(parent(x),x)* by *dif=(l'(parent(x),x)-lmin(parent(x),x))*, we increase $C_{total}$ by *dif* and we range decrease the values in the interval *I(x)* of the segment tree *A* by *dif*; afterwards, we set *l'(parent(x),x)* to *lmin(parent(x),x)* and we query the distance *CD* again (from the leaf *DFSnum(i)* of the segment tree *A*). If, after performing these actions, we have *l'(parent(x),x)=lmin(parent(x),x)*, then: for every son *s(x,j)* of vertex *x* we range set all the values in the interval *I(s(x,j))* of the segment tree *B* to *s(x,j)*). If, at some point, *CD>D* and the latency of the edge *(parent(x),x)* cannot be decreased at all *(l'(parent(x),x)* is already equal to *lmin(parent(x),x)*; *dif=0)* and *x* is a leaf in the tree, then the candidate distance *D* is not feasible. If, at the end, $C_{total}$ is larger than *C*, *D* is not feasible. If *D* is not feasible, we will test a larger candidate distance in the binary search; otherwise, we will test a smaller one. The time complexity of this approach is $O(n \cdot log(n) \cdot log(DMAX))$.

## 4. NETWORK DESIGN PROBLEMS

For the first problem, we are given a complete graph (network) with *n* vertices. Every edge *(u,v)* has a label *l(u,v)* (between *1* and *q*). We want to obtain a spanning subgraph *H* of the complete graph, such that the distance between any two vertices in *H* is at most three (the shortest path between any two vertices contains at most three edges) and the number of distinct labels of the chosen edges is as small as possible. For this problem we will present a greedy, heuristic algorithm. We want the obtained network to have the following structure: a central edge *(x,y)* to which every other vertex is connected (i.e. every vertex *z≠x* and *z≠y* is connected either to *x* or to *y*). Obviously, such a network has diameter at most three. Let's assume that the central edge is fixed. We will now traverse the remaining *n-2* vertices in an arbitrary order *v(1), …, v(n-2)* (or we can use a heuristic algorithm to choose the order). We will maintain an array *used*, where *used(a)=true* if label *a* has already been used. Initially, we have *used(l(x,y))=true* (and *used(e≠l(x,y))=false*). For each vertex *v(i)* (*i=1,…,n-2*) we first test if either *used(l(x,v(i)))=true* (in which case we connect *v(i)* to *x*) or *used(l(y,v(i)))=true* (in which case we connect *v(i)* to *y*). If both labels *(l(x,v(i))* and *l(y,v(i)))* were not used, yet, we will need to choose one of the labels. If *l(x,v(i))=l(y,v(i))*, then there is no choice to make: we connect *v(i)* to *x* and set *used(l(x,v(i)))* to *true*. Otherwise, we will compute *nx* (*ny*), the number of vertices *v(j)* (*i≤j≤n-2*) such that at least one of the labels *l(x,v(j))* and *l(y,v(j))* is equal to *l(x,v(i))* (*l(y,v(i))*) and *used(l(x,v(j)))=used(l(y,v(j)))=false*. If *nx≥ny*, we will connect *v(i)* to *x* and set *used(l(x,v(i)))* to *true*; otherwise, we connect *v(i)* to *y* and set *used(l(y,v(i)))* to *true*. The time complexity of this algorithm is $O(n^2)$ (if the edge *(x,y)* is fixed). If the total number of distinct labels (*q*) is not too large, we can compute in the beginning the values *num(j)*=the number of vertices *v(i)* such that: *(l(x,v(i))=j)* or *(l(y,v(i))=j)* (or both) ($1 \leq j \leq q$). We initialize *num(j)* to *0* ($1 \leq j \leq q$) and then we traverse the vertices *v(i)* (but we skip over those vertices for which *l(x,v(i))* or *l(y,v(i))* are equal to *l(x,y)*); if *l(x,v(i))≠l(y,v(i))*, we increment by *1* both *num(l(x,v(i)))* and *num(l(y,v(i)))*; otherwise, we only increment *num(l(x,v(i)))* by *1*. We also maintain *q* lists *Li(1), …, Li(q)*. We insert every (non-skipped) vertex *v(k)* into *Li(l(x,v(k)))* and, if *l(x,v(k))≠l(y,v(k))*, also into *Li(l(y,v(k)))*. After this, we run the actual algorithm. Whenever we need to compute *nx* and *ny* for a vertex *v(i)*, we have *nx= num(l(x,v(i)))* and *ny=num(l(y,v(i)))*. Whenever we set *used(j)=true* (where *used(j)* was previously equal to *false*), we traverse the list *Li(j)* and, for each vertex *v(k)* in *Li(j)*, we remove it from *Li(j)* and from any other list into which it is contained (*v(k)* may be contained in at most *2* lists). Whenever we remove a vertex *v(k)* from a list *Li(p)* (into which it was previously contained), we decrease *num(p)* by *1*. The time complexity of this approach is *O(n+q)*. By noticing that there can be at most *q'=2·(n-2)=O(n)* distinct labels on the edges adjacent to *x* or *y* (which can be renumbered from *1* to *q'*), the time complexity becomes *O(n+q')=O(n)* in any case. In order to complete the algorithm, we need to test several possibilities for the edge *(x,y)*. The best approach would be to consider every edge *(u,v)* as a candidate edge *(x,y)* and run the algorithm for every edge (the time complexity would be $O(n^4)$ or $O(n^3)$). If the time complexity is too high, we can choose the vertex *x* (arbitrarily or according to some other heuristic, e.g. the vertex which is adjacent to edges whose set of labels contains the smallest total number of distinct labels) and consider every edge *(x,v)* as a candidate edge *(x,y)* (this reduces the time complexity by a factor of *O(n)*, obtaining an $O(n^2)$ time complexity).

In the second problem, for reliability purposes, we want to construct a connected graph with *n* vertices, where the degree of each vertex is exactly *k* (a *k*-regular graph). We will first present a solution for even *k* and then we will present a general solution. In order to generate a *k*-regular graph we will start from a complete graph having *k+1* nodes (which is, obviously, *k*-regular) and we will add one node at a time, forming a new *k*-regular graph. We will be interested only in the nodes *1,2,...,k* which are divided into *2* groups *{1,2,...,k/2}* and *{k/2+1,...,k}*. We will ignore the edges between two nodes of the same group. Under these circumstances, we will only look at the complete bipartite sub-graph which has nodes *1,2,...,k/2* on the left side and *k/2+1, k/2+2,..., k* on the right side. We will insert node *k+2* at the middle of the edges *(1,(k/2+1)), (2,(k/2+2)), ..., ((k/2),k)*. Inserting a node *a* at the middle of an edge *(b,c)* introduces the new edges *(b,a)* and *(a,c)*, but removes the edge *(b,c)*. We notice that all the nodes from *1* to *k+1* maintain their degree *k* and the newly inserted node also has degree *k*. Inserting another node *(k+3)* is performed similarly, but the replaced edges will be *(1,(k/2+2)), (2,(k/2+3)), ..., ((k/2-1),k), ((k/2),(k/2+1))*. What is important is that a complete matching of the previously mentioned bipartite graph can be selected. We can easily find a way to choose edges so that *k/2* nodes are inserted (i.e. *k/2* distinct matchings are selected). For instance, when inserting the node *(k+1+x)* (*1≤x≤k/2*), the *k/2* replaced edges will be *(i, k/2+((i+x-2) mod (k/2))+1)* (*1≤i≤k/2*). After inserting *k/2* nodes, we notice that we would obtain a new bipartite complete sub-graph if we would consider the nodes *{1,2,..,k/2}* and the *k/2* newly inserted nodes. Using this bipartite sub-graph, we can insert another set of (up to) *k/2* nodes, and so on (until we insert *(n div (k/2))* complete sets of *k/2* nodes, plus *(n mod (k/2))* final nodes). The time complexity is $O(n \cdot k)$. The general solution uses a well-known algorithm of decomposing the edge set of a complete graph into *((n-1) div 2)* disjoint Hamiltonian cycles (and a 1-factor, if *n* is even). We choose any *(k div 2)* Hamiltonian cycles of the decomposition. If *k* is odd, we also choose the 1-factor.

## 5. NETWORK CLUSTERING IN PATH NETWORKS

In this section we present efficient algorithms for several constrained and unconstrained clustering problems in path networks. These problems are better expressed in geometric terms. We consider *n* points located on the real line, given in increasing order of their x-coordinates: $x(1) \leq x(2) \leq ... \leq x(n)$. Each point *i* is located at coordinate *x(i)* and has *T* (or $k \cdot T$) non-negative weights: *w(i,[j,]1), ..., w(i,[j,]T)* (*1≤j≤k*). We want to split the points into several disjoint intervals (clusters), such that the value of an objective function (*objf*) is minimized. A cluster *[a,b]* contains all the points *i* with *a≤i≤b*. The objective function will be an aggregate (*sum, max*) over the costs of the clusters. For each cluster type *tc* (*1≤tc≤T*) we have an aggregate function *ctype(tc)* which aggregates the weights *w(i,[j,]tc)* of the points *i* in the cluster. For a given cluster *c*, let *tcagg(c,tc)* be the result of the function *ctype(tc)*, applied to all the points in *c*. Then, the cost function of each cluster *c* will be an aggregate (*ccost*) over the *tcagg(c,tc)* values (any function with *T* parameters is correct), plus a fixed value *F*. The clustering constraints will be given as the number of clusters (*1≤k≤n*) and/or as some values *1≤l(i,[j,]tc)≤u(i,[j,]tc)≤i*, denoting the smallest index of a point which can be included in the same cluster as point *i* (*l(i,[j,]tc)*) and the smallest index of a point which must necessarily be included in the same cluster as point *i* (*u(i,[j,]tc)*), if the cluster's type is *tc* and point *i* is the rightmost point in the cluster (its representative). These values may be given implicitly, by stating, for instance, that each point *i* may be the representative (rightmost point) of a cluster of type *tc* of length at most *lmax(i,[j,]tc)* and at least *lmin(i,[j,]tc)*, or that the total weight of the points inside a cluster of type *tc* whose representative is point *i* is at most *wmax(i,[j,]tc)* and at least *wmin(i,[j,]tc)*. If given implicitly, we can compute all the *l(i,[j,]tc)* and *u(i,[j,]tc)* values in $O(n \cdot T)$ time by sweeping the points (when *lmin(i,[j,]tc)≤ lmin(i-1,[j,]tc)+|x(i)-x(i-1)|* and *lmax(i,[j,]tc)≤lmax(i-1,[j,]tc)+|x(i)-x(i-1)|*, or *wmax(i,[j,]tc)≤wmax(i-1,[j,]tc)+w(i,[j,]tc)* and *wmin(i,[j,]tc)≤wmin(i-1,[j,]tc)+w(i,tc)* for all *2≤i≤n*), or by binary searching and prefix weight-sum computations (for arbitrary values of *lmin(i, [j,] tc), lmax(i, [j,] tc), wmin(i, [j,] tc)* or *wmax(i, [j,] tc)*). In the first case, once we computed *l(i-1, [j,] tc)* (*u(i-1, [j,] tc)*), we can compute *l(i, [j,] tc)* (*u(i, [j,] tc)*) by initializing it to *l(i-1, [j,] tc)* (*u(i-1, [j,] tc)*) and increasing it by *1* until we reach *i* or the first point *l(i, [j,] tc)* (last point *u(i, [j,] tc)*) for which the distance between point *i* and this point is at most *lmax(i, [j,] tc)* (at least *lmin(i, [j,] tc)*) (or for which the sum of the weights *w(*,[j,]tc)* of the points in the interval *[l(i,[j,]tc),i]* (*[u(i,[j,]tc),i]*) is at most *wmax(i,[j,]tc)* (at least *wmin(i,[j,]tc)*)). In the second case, we binary search *l(i,[j,]tc)* (*u(i,[j,]tc)*) in the interval *[1,i]*, because we have the property that for all the points *p* from *1* to *l(i,[j,]tc)-1* (*u(i,[j,]tc)*), the distance up to point *i* [sum of the weights *w(*,[j,]tc)* of the points in the interval *[p,i]*] is larger than *lmax(i,[j,]tc)* [*wmax(i,[j,]tc)*] (larger than or equal to *lmin(i,[j,]tc)* [*wmin(i,[j,]tc)*]), and for *p≥l(i,[j,]tc)* (*u(i,[j,]tc)+1*), the distance up to *i* [sum of weights of the points in the interval *[p,i]*] is smaller than or equal to *lmax(i, [j,]tc)* [*wmax(i,[j,]tc)*] (smaller than *lmin(i, [j,]tc)* [*wmin(i, [j,]tc)*]). The middle argument *j* of any value *val(i,j,tc)* will always refer to the case when point *i* belongs to the *j*[th] cluster, counting from left to right, and the number of clusters *k* is given (if *k* is not given, we will have *val(i,tc)*, instead of *val(i,j,tc)*). When the number of clusters is fixed (*k*) we will compute $C_{min}(i,j)$=the minimum value of the objective function, if the points *1,2,...,i* are split into *j* clusters. We have $C_{min}(0,0)=0$ and $C_{min}(i>0,0)= C_{min}(0,j>0)=+\infty$. For *i≥j>0*, we will initialize $C_{min}(i,j)=+\infty$ and then we will consider every point *p* as the first point of the *j*[th] cluster (the last point is point *i*), in decreasing order (starting from *p=i* and ending at *p=j*). We will maintain the values *tcagg(tc)* of the weights of the points in the interval *[p,i]* (initially, these values will be *undefined*). When we reach a new value of *p*, we update the values *tcagg(tc)* (*1≤tc≤T*): *tcagg(tc)= ctype(tc)(w(p,[j,]tc), tcagg(tc))* (if the previous value *tcagg(tc)=undefined*, then *tcagg(tc)* will be equal to *w(p, [j,]tc)*). If *p<l(i,[j,]tc)* or *p>u(i,[j,]tc)*, we will set *tcagg2(tc)* at a value which shows that a constraint is violated (e.g. we set *tcagg2(tc)=+∞*), i.e. a value which will increase the value of the functions *ccost* and *objf* very much; otherwise, we set *tcagg2(tc)=tcagg(tc)*. Then, we recompute the aggregate cost of the cluster: *cc=ccost(tcagg2(1), …, tcagg2(T))*. We set $C_{min}(i,j)=min\{C_{min}(i,j), objf(C_{min}(p-1, j-1), F+cc)\}$. This algorithm has an $O(n^2 \cdot k \cdot T)$ time complexity. When the number of clusters *k* is not given, we can drop the second index (*j*) from the state of the dynamic programming (DP) algorithm and compute $C_{min}(i)$=the minimum value of *objf*, if the points *1,2,...,i* are split into any number of clusters ($C_{min}(0)=0$ and $C_{min}(i>0)=+\infty$, initially; $C_{min}(i)=min\{C_{min}(i), objf(C_{min}(p-1), F+cc)\}$). In this case, the time complexity is $O(n^2 \cdot T)$. In the rest of the section we will only be interested in the case *ccost=min* and *ctype(tc)=ctype()=sum* or *max*, i.e. the cost of a cluster is the minimum of the costs of each cluster type and we use the same aggregation function to compute the cost for each cluster type *tc* (and this function is either *sum* or *max*). This is the same as choosing the most convenient type of cluster. In the unconstrained case, for *objf=max, ctype=sum* or *max*, the optimal solution consists of *n* clusters: *[1,1], ..., [n,n]*; the type *tc* of each cluster *i* is the one for which $w(i,tc)=min\{w(i,t')|1 \leq t' \leq T\}$. We will present next significant improvements for the constrained cases for each of the *4* pairs (*objf, ctype*). Every time we will ask for the *min* (*max*) element (field of a tuple) of an empty set or data structure, the result will be $+\infty$ ($-\infty$). All the used data structures are emptied after computing all the values $C_{min}(*,j)$ (for every $1 \leq j \leq k$, when *k* is given); when *k* is not given, the data structures are only emptied once, in the beginning.

### 5.1. *objf=sum, ctype=sum*

For each cluster type *tc* (and every cluster index *j+1*) we will compute in $O(n)$ time the prefix sums $wp(i,[j+1,]tc) = w(1,[j+1,]tc)+...+w(i,[j+1,]tc) = wp(i-1,[j+1,]tc)+w(i,[j+1,]tc)$ ($wp(0,[j+1,]tc)=0$). When we have $l(*,[*,] *)=1$ and the number of clusters *k* is given, we will use the following strategy. Let's assume that all the values $C_{min}(*,j)$ were computed and we are ready to begin computing the values $C_{min}(i,j+1)$ (in increasing order of *i=1,...,n*). While doing this, we will compute a table $D_j(i,tc)$. We have $D_j(0,tc)=C_{min}(0,j)$ and $D_j(i>0,tc)=min\{D_j(i-1,tc), C_{min}(i,j)-wp(i,[j+1,]tc)\}$. We compute $D_j(i-1,tc)$ just before computing $C_{min}(i,j+1)$. Then, with this table, we can compute $C_{min}(i,j+1)$ in $O(T)$ time, as $min\{D_j(u(i,[j+1,]tc)-1,tc)+wp(i,[j+1,]tc)+F|1 \leq tc \leq T\}$. Thus, the complexity becomes $O(n \cdot k \cdot T)$. When the constraints $l(i,[j,]tc)$ are arbitrary, we can use a similar approach. After computing all the values $C_{min}(*,j)$ for a given *j*, we will compute a table $E_j(*,tc)$, with $E_j(i \geq 0,tc)=C_{min}(i,j)-wp(i,[j+1,]tc)$. $C_{min}(i,j+1)=F+min\{wp(i,[j+1,]tc)+min\{E_j(p,tc)|l(i,[j+1,]tc)-1 \leq p \leq u(i,[j+1,]tc)-1\}|1 \leq tc \leq T\}$. By building a segment tree over each column $E_j(*,tc)$ of each table $E_j$, we can find the minimum value in any interval of rows of any column in $O(log(n))$ time, improving the overall complexity to $O(n \cdot log(n) \cdot k \cdot T)$. The segment tree also supports updates, s.t. we can initialize the $E_j(*,tc)$ to $+\infty$ and set $E_j(i-1,tc)$ to the correct value right before computing $C_{min}(i,j+1)$. An alternative is to construct the whole table $E_j(*,*)$ and then preprocess it, in order to answer range minimum queries on each column - the complexity may drop by an $O(log(n))$ factor. Further improvements are possible if $l(a,[j,]tc) \leq l(b,[j,]tc)$ and $u(a,[j,]tc) \leq u(b,[j,]tc)$ for every cluster type *tc*, any $1 \leq j \leq k$, and any two points *a<b* (the *non-decreasing property*). When computing the values $C_{min}(i,j+1)$, we will maintain an array *DQ* of *T* double-ended queues (deques); each deque contains (*index, value*) pairs. We initialize *DQ[tc]* ($1 \leq tc \leq T$) by inserting the pair ($-1,+\infty$). Then, we begin computing the values $C_{min}(i,j+1)$ in increasing order of *i*. For a given *i*, and every cluster type *tc*, we insert in increasing order of *p* at the end of each deque *DQ[tc]*, the pairs $pr(i,[j+1],tc,p)=(index=p, value=C_{min}(p,j)-wp(p,[j+1,]tc))$, with $u(i-1,[j+1,]tc) \leq p \leq u(i,[j+1,]tc)-1$ (we consider $u(0,[*,]*)=0$). Before inserting a pair (*idx,val*) into a deque, we repeatedly remove from the end of the deque the last pair, as long as its *value* field is larger than *val* (and the deque is not empty). Afterwards, we repeatedly remove from the beginning of each deque *DQ[tc]* the first pair, as long as its *index* field is smaller than $l(i,[j+1,]tc)-1$. After all these operations, we compute (in $O(T)$ time) the value $V=min\{DQ[tc].getFirst().value+wp(i,[j+1,]tc)|1 \leq tc \leq T\}$ and set $C_{min}(i,j+1)$ to $V+F$. The total number of insertions (deletions) into (from) each deque is $O(n)$ and each such operation is performed in $O(1)$ (amortized) time. The time complexity of the algorithm is $O(n \cdot k \cdot T)$. Instead of deques, we could have used two arrays of min-heaps, $H_{idx}[*]$ and $H_{val}[*]$: before computing $C_{min}(i,j+1)$, for every cluster type *tc*, we: (1) insert all the pairs $pr(i,[j+1],tc,p)$ into both $H_{val}[tc]$ and $H_{idx}[tc]$; (2) while $H_{idx}[tc].getMinIndex()$ is smaller than $l(i,[j+1,]tc)-1$, we remove the tuple with the smallest *index* field both from $H_{idx}[tc]$ and $H_{val}[tc]$. $C_{min}(i,j+1)=min\{H_{val}[tc].getMinValue()|1 \leq tc \leq T\}+wp(i,[j+1,]tc)+F$. In this case, the time complexity is $O(n \cdot log(n) \cdot k \cdot T)$. When the number of clusters *k* is not given, we can modify the solutions presented above by dropping the index *j* (the number of clusters) from the state of the DP (except for the RMQ approach for the tables $E_j$). We compute $C_{min}(i)$ instead of $C_{min}(i,*)$, we replace $C_{min}(i-1,*)$ by $C_{min}(i-1)$ and we drop the index *j* from the definitions of the tables *D* and *E*. The tables *D* and *E*, the deques (or heaps) are updated during the single traversal of the points *i=1,...,n*. The complexities mentioned before are decreased by a factor of *k*. These techniques work even with negative weights.

### 5.2. *objf=sum, ctype=max* and $u(i,[*,]*)=i$

When all the $l(i,[j,]tc)$ values are *1* and the number of clusters *k* is given, we will maintain an array *S* of *T* stacks, each stack containing (*index, vmax, pcmin, smin*) tuples. Before computing a value $C_{min}(i,j+1)$ (in increasing order of *i=1,...,n*), we perform the following computations for each cluster type *tc*: (1) we build a tuple $tu(i,[j+1,]tc)=(index=i, vmax=w(i,[j+1,]tc), pcmin=C_{min}(i-1,j), smin=w(i,[j+1,]tc)+C_{min}(i-1,j))$; (2) while the topmost tuple *tp* in *S[tc]* has $tp.vmax \leq tu(i,[j+1,]tc).vmax$, we pop *tp* from the stack, set $tu(i,[j+1,]tc).pcmin$ to $min\{tu(i,[j+1,]tc).pcmin, tp.pcmin\}$ and, after this, we set $tu(i,[j+1,]tc).smin$ to $w(i,[j+1,]tc)+tu(i,[j+1,]tc).pcmin$; (3) if *S[tc]* is not empty, let *tp* be the topmost tuple in *S[tc]*: we set $tu(i,[j+1,]tc).smin$ to $min\{tu(i,[j+1,]tc).smin, tp.smin\}$; (4) we push $tu(i,[j+1,]tc)$ on *S[tc]*. We will set $C_{min}(i,j+1)$ to $F+min\{S[tc].getTopmostTuple().smin|1 \leq tc \leq T\}$. The time complexity is $O(n \cdot k \cdot T)$. When $l(i,[j,]tc) \leq l(i+1,[j,]tc)$ ($1 \leq i \leq n-1; 1 \leq j \leq k; 1 \leq tc \leq T$), we can use an array of deques *DQ* (instead of an array of stacks). Each deque *DQ[tc]* stores (*index, vmax, pcmin*) tuples (we dropped the *smin* field). Then, before computing the value $C_{min}(i,j+1)$, we perform the same operations as in the previous solution, with the following differences: the tuple $tu(i,[j+1,]tc)$ does not have the *smin* field; the top of the stack *S[tc]* now becomes the end of the deque *DQ[tc]*; popping a tuple from the stack=removing the last tuple in the deque; pushing a tuple on the stack *S[tc]*=inserting a tuple at the end of the deque *DQ[tc]*; any operation referencing the field *smin* is dropped. We perform the following extra action: as long as the *index* field of the tuple located at the beginning of the deque *DQ[tc]* ($1 \leq tc \leq T$) is smaller than $l(i,[j+1,]tc)$, we remove the tuple from *DQ[tc]*. If *DQ[tc]* is not empty, we set the *pcmin* field of the first tuple (at the front) of *DQ[tc]* to $min\{C_{min}(p,j)|l(i,[j+1,]tc)-1 \leq p \leq DQ[tc].getFirst().index-1\}$; in order to compute this minimum value in $O(log(n))$ time, we can construct a segment tree $ST_j$ over the values $C_{min}(*,j)$ and set the value of the leaf *i-1* of $ST_j$ to $C_{min}(i-1,j)$ right before computing $C_{min}(i,j+1)$; we can also preprocess all the values $C_{min}(*,j)$, in order to answer RMQ queries in $O(1)$ time; or we can use deques to maintain the minimum in a window whose endpoints, $l(i,[j+1,]tc)-1$ and $DQ[tc].getFirst().index-1$, only increase, in $O(1)$ amortized time. After this, we need to compute the value $V=min\{tp.vmax+tp.pcmin|tp \in DQ[tc], 1 \leq tc \leq T\}$ and set $C_{min}(i,j+1)$ to $F+V$. We will maintain all the tuples *tp* inside all the deques in a min-heap *H*, where their key is ($tp.vmax+ tp.pcmin$). Whenever we remove a tuple from a deque, insert a new tuple inside a deque or change the *vmax* or *pcmin* fields of a tuple in a deque, we also update the heap *H* (by *inserting/removing/changing the key of* the tuple *in/from/in H*). The overall time complexity is $O(n \cdot log(n \cdot T) \cdot k \cdot T)$ (if we maintain a different min-heap *HP[tc]* for every cluster type *tc* and compute each value $C_{min}(i,*)$ in $O(T)$ time, the time complexity would be $O(n \cdot log(n) \cdot k \cdot T)$). When *k* is not given, we can drop the index *j (j+1)* from the DP state and replace every reference $C_{min}(i,*)$ by $C_{min}(i)$; the RMQ approach cannot be extended to this case this time, either.

*5.3. objf=max, ctype=sum*

An easy solution when all the values $l(*,[*,]*)$ are *1*, $u(i,[*,]*)=i$ and $w(i,[*,]tc)$ are equal (i.e. $w(i,[1,]tc)=...=w(i,[k,]tc)$; we will denote these values by $w(i,tc)$, as the middle argument *j* makes no difference), would be to binary search the optimal value *OPT* of the objective function. Let's assume that we want to test the value $O_{cand}$, selected by the binary search. We can do this by traversing the points from *1* to *n* and maintaining several counters: *nc*, representing the number of clusters (initialized to *1*), and *tsum[tc]* (initialized to *0*). Then, for each point *i* and each cluster type *tc*, we add $w(i,tc)$ to *tsum[tc]*; if all the counters *tsum[tc]* become larger than $O_{cand}$-*F*, we increment *nc* by *1* and set the values of each counter *tsum[tc]* to $w(i,tc)$. If $w(i,tc)>O_{cand}$-*F* ($1 \leq tc \leq T$), then we will always have *tsum[tc]*=+∞ from now on (for the next points *i+1,...,n*). If, at some point, all the values *tsum[\*]* are +∞, then $O_{cand}$ is not feasible (we set *nc=k+1*). This test minimizes the number of clusters, such that the sum of the weights of the points in each cluster is at most $O_{cand}$-*F*. If (*nc>k*), then $O_{cand}$ is not feasible. If $O_{cand}$ is feasible, we can test a smaller value than $O_{cand}$ next; otherwise, we will test a larger one. It is obvious that if we can split the *n* points into *nc<k* clusters such that the value of the objective function is at most $O_{cand}$, we can always split further some of the clusters and form exactly *k* clusters, without increasing the value of the objective function. The time complexity of this approach is $O(n \cdot T \cdot log(WMAX))$, where *WMAX* is the sum of the largest weights of the points. If the weights are integers, we always find the optimal solution; if they are real numbers, we can only approximate the optimal answer with any accuracy $\varepsilon>0$. An exact solution for the more general case with $l(*,[*,]*)=1$, $u(i,[j,]tc) \leq u(i+1,[j,]tc)$ ($1 \leq i \leq n-1$; $1 \leq j \leq k$; $1 \leq tc \leq T$) and not necessarily equal $w(i,[*,]tc)$ values, is the following. We compute the prefix sums $wp(i,[j+1,]tc)$ (defined previously). Then, after computing a value $C_{min}(i,j)$, we compute the indices $r(i,tc,j)$ of the rightmost points, such that the condition $wp(r(i,tc,j), [j+1,] tc)-wp(i, [j+1,] tc) \leq C_{min}(i,j)$ holds. We can compute every index in $O(log(n))$ time, by using binary search. However, in this case, we can do better. Because $C_{min}(i+1,j) \geq C_{min}(i,j)$ (for $i=pf,...,n-1$, where *pf* is the smallest index with $C_{min}(pf,j)<+\infty$) and $wp(i+1,[j+1,]tc) \geq wp(i,[j+1,]tc)$, we have that $r(i+1,tc,j) \geq r(i,tc,j)$ ($pf \leq i \leq n-1$). Thus, we can search for $r(i+1,tc,j)$ by initializing it to $r(i,tc,j)$ (or to *1*, if $C_{min}(i,j)=+\infty$, in which case $r(i,tc,j)=n$) and repeatedly increasing it by *1*, as long as the condition holds. Thus, $O(n)$ time is spent overall for computing all the values $r(*,tc,j)$ ($1 \leq tc \leq T$). While computing the values $C_{min}(*,j+1)$, we will maintain an array *DQ* of *T* deques; each deque *DQ[tc]* contains (*index, limit, value*) tuples. We will also maintain an array *smax* of *T* values; each value *smax[tc]* is initialized to -∞. Before computing a value $C_{min}(i,j+1)$ ($i=1,...,n$), for each cluster type *tc* we perform the following actions: (1) we insert the tuples $tu(i,p)=(index=p, limit=r(p,tc,j), value=C_{min}(p,j))$, with $u(i-1,[j+1,]tc) \leq p \leq u(i,[j+1,]tc)-1$ ($u(0,[*,]*)=0$), in increasing order of *p*, at the end of the deque *DQ[tc]* (we only insert a tuple $tu(i,p)$ after repeatedly removing the last tuple *tlp* from *DQ[tc]*, while $tlp.value \geq tu(i,p).value$); (2) while the first tuple *tp* of the deque *DQ[tc]* has the *limit* field smaller than *i*, we remove *tp* from *DQ[tc]* and set $smax[tc]=max\{smax[tc], wp(tp.index,[j+1,]tc)\}$. Then, we compute $V=min\{DQ[tc].getFirst().value|1 \leq tc \leq T, DQ[tc]$ is not empty$\}$ and $U=min\{wp(i,[j+1,]tc)-smax[tc]|1 \leq tc \leq T\}$ and set $C_{min}(i,j+1)$ to $min\{U,V\}+F$. *U* is the minimum sum of the weights of the points in the cluster containing point *i*, with the property that this sum is larger than the sum of all the previous clusters. *V* is the minimum cost of a previous cluster, with the property that this cost is the largest among all the chosen clusters (including point *i*'s cluster). The complexity is $O(n \cdot k \cdot T)$. The case with *non-decreasing* values $l(i,[j,]tc)$ and $u(i,[j,]tc)$ can be solved by adapting the solution mentioned above. We transform the *smax* array into an array of deques; each deque contains (*index, value*) pairs. Then, before computing $C_{min}(i,j+1)$, we remove the first tuples of *DQ[tc]* and *smax[tc]*, as long as their *index* field is smaller than $l(i,[j+1,]tc)-1$ (for $1 \leq tc \leq T$). Afterwards, while the *limit* field of the first tuple *tp* of *DQ[tc]* ($1 \leq tc \leq T$) is smaller than *i*, we remove it from *DQ[tc]* and insert $tnew=(index=tp.index, value=wp(tp.index, [j+1,]tc))$ at the end of *smax[tc]*; before doing this, we repeatedly remove the tuple *tlast* from the end of *smax[tc]*, as long as *smax[tc]* is not empty and $tlast.value \leq tnew.value$. Then, the value *U* is defined as $min\{wp(i,[j+1,]tc)-smax[tc].getFirst().value|1 \leq tc \leq T, smax[tc]$ is not empty$\}$. The complexity for this case is $O(n \cdot k \cdot T)$, too. The case with arbitrary $l(i,[j,]tc)$ and $u(i,[j,]tc)$ values is handled differently. While computing the values $C_{min}(*,j+1)$, we will maintain two arrays of $T$ 2D range trees, $A$ and $B$. In each tree we insert $n+1$ dummy points $(i,-\infty)$ with weights $+\infty$ for $A[tc]$, and $-\infty$ for $B[tc]$ $(i=0, \ldots, n)$. A 2D range tree is a segment tree, where each leaf stores a point; the points are sorted according to the x-coordinate from the leftmost leaf to the rightmost one. Each internal node stores all the points contained in the leaves of its subtree; thus, every point is stored in $O(log(n))$ tree nodes. Each tree node $q$ of the range tree stores all of its points in an augmented balanced tree $T_q$ (e.g. AVL tree, red-black tree, scapegoat tree). The points are inserted in $T_q$ with their y-coordinate as the key. Each node $q'$ in $T_q$ also maintains the smallest weight $minw(q')$ of a node in its subtree (in $T_q$). We can insert or delete a pair $(y, weight)$ in $T_q$ and maintain the values $minw$ in logarithmic time. We can also search for the smallest weight within an interval $[y_1, y_2]$ with the same complexity, if we additionally maintain in every node of $T_q$ the smallest and largest y-coordinates of a point in its subtree. The function $findMinW(x_1, y_1, x_2, y_2)$ of a range tree returns the minimum weight of a point $(x, y)$ in the tree, with $x_1 \leq x \leq x_2$ and $y_1 \leq y \leq y_2$ (or $+\infty$ if no point lies in the range), in time $O(log^2(n))$. Similarly, we can support a $findMaxW(x_1, y_1, x_2, y_2)$ function for a range tree, which returns the maximum weight of a point in the given range (or $-\infty$ if no point exists in the range). We will compute the $r(*,*,*)$ values by binary searching each of them. Before computing $C_{min}(i, j+1)$ $(i=1, \ldots, n)$, for each cluster type $tc$, as before, we consider the values $p$, $u(i-1,[j+1,]tc) \leq p \leq u(i,[j+1,]tc)-1$; for each value, if the dummy points $(p,-\infty)$ exist in $A[tc]$ and $B[tc]$, we remove them from there and we insert the point $(p, r(p,tc,j))$ with the weight $C_{min}(p,j)$ into the range tree $A[tc]$ and the point $(p, r(p,tc,j))$ with the weight $wp(p,[j+1,]tc)$ into the range tree $B[tc]$ $(1 \leq tc \leq T)$. We define $V=min\{A[tc].findMinW(l(i,[j+1,]\ tc)-1,i,u(i,[j+1,]tc)-1,+\infty) \mid 1 \leq tc \leq T\}$ and $U=min\{wp(i,[j+1,]\ tc)-B[tc].findMaxW(l(i,[j+1,]tc)-1,-\infty,u(i,[j+1,]tc)-1,i-1) \mid 1 \leq tc \leq T\}$ and, like before, we set $C_{min}(i,j+1)=min\{U,V\}+F$. When $k$ is not given, we modify the above solutions by dropping the index $j$ (the cluster's index) from the DP state and from all equations and by traversing the points from $1$ to $n$ only once.

### 5.4. objf=max, ctype=max

When the number of clusters $k$ is given, all the values $l(*,[*,]*)$ are $1$, $u(i,[*,]*)=i$ and $w(i[,*],tc)$ are equal, we can use the binary search approach in this case, too. The difference consists of replacing the $tsum[tc]$ counters by the $tmax[tc]$ values $(1 \leq tc \leq T)$. We initialize $tmax[*]$ to $-\infty$. Then, we traverse the points from $i=1$ to $n$ and set $tmax[tc]=max\{tmax[tc], w(i,tc)\}$. When all the values $tmax[*]$ exceed $O_{cand}-F$ after considering point $i$, we increase the number of clusters $nc$ and reset $tmax[tc]$ to $w(i,tc)$ $(1 \leq tc \leq T)$. If $w(i,tc)> O_{cand}-F$ $(1 \leq tc \leq T)$, then we will always have $tmax[tc]=+\infty$ from now on (for the next points $i+1, \ldots, n$). If, at some point, all the values $tmax[*]$ exceed $O_{cand}-F$, then $O_{cand}$ is not feasible. If the candidate values $O_{cand}$ are taken from the sorted list of $n \cdot T$ point weights (increased by $F$), the answer is exact and the time complexity is $O(n \cdot T \cdot log(n \cdot T))$. When $l(*,[*,]*)=1$, $u(i,[j,]tc) \leq u(i+1,[j,]tc)$ $(1 \leq i \leq n-1; 1 \leq j \leq k; 1 \leq tc \leq T)$ and we have no constraints on the non-negative values $w(*,[*,]*)$, we will compute again the values $r(i,tc,j)$=the largest index of a point such that $max\{w(i+1,[j+1,]tc), \ldots, w(r(i,tc,j),[j+1,]tc)\} \leq C_{min}(i,j)$. In this case, for fixed $tc$ and $j$, these values are non-decreasing (as $i$ increases, starting from the smallest value $pf \geq 0$ with $C_{min}(pf,j)<+\infty$; for $0 \leq i \leq pf-1$, $r(i,tc,j)=n$) and can be computed in $O(n)$ time (otherwise, we can binary search them). We can compute $RMQwmax(a,b,[j,]tc)$=the maximum weight $w(q,[j,]tc)$, with $a \leq q \leq b$, in $O(1)$ time using RMQ (Bender et al., 2000) (with $O(n[\cdot k])$ or $O(n[\cdot k]\cdot log(n))$ preprocessing) for each cluster type $tc$. We adapt the solutions from the previous case. For non-decreasing $l(i,[j,]tc)$ and $u(i,[j,]tc)$ values, we will use the same $smax[tc]$ values as in the case $l(*,[*,]*)=1$. Whenever we remove a tuple $tp=(index=idx, limit=lim, value=val)$ from the front of $DQ[tc]$ because $lim<i$, we set $smax[tc]=max\{smax[tc], tp.index\}$. Then $U$ is defined as $min\{RMQwmax(smax[tc]+1,i,[j+1,]tc) \mid 1 \leq tc \leq T, smax[tc] \geq l(i,[j+1,]tc)-1\}$. For arbitrary $l(i,[j+1,]tc)$ and $u(i,[j+1,]tc)$ values, we compute the $r(*,*,*)$ values by binary searching each of them. The weights of the points $(p, r(p,tc,j))$ inserted in $B[tc]$ will be $p$, and $U$ is defined as $min\{RMQwmax(B[tc].findMaxW(l(i,[j+1,]tc)-1,-\infty,u(i,[j+1,]tc)-1,i-1)+1,i,[j+1,]tc)$ (or $+\infty$, if $B[tc].findMaxW(l(i,[j+1,]tc)-1,-\infty, u(i,[j+1,]tc)-1, i-1)<0) \mid 1 \leq tc \leq T\}$. The case when $k$ is not given is handled by transforming the solutions mentioned above, just like in all the other cases.

### 5.5. Clustering of Mobile Network Devices

The final clustering problem that we consider consists of a network of mobile devices. Each device $i$ $(1 \leq i \leq n)$ is a point on the real line initially located at position $x(i)$ and moves in direction $d(i)$ ($d(i)=-1$ for *left* or $+1$ for *right*) at a speed of $v(i)$ distance units per time unit. We want to find the earliest time moment when we can place $K$ identical intervals of fixed given length $L$, such that every point is inside one of the intervals (if possible). The motivation is given by the fact that we want to send a piece of content to $K$ devices. Each device can send data within an interval of length $L$ containing its location. Then, when all the devices can be contained within $K$ intervals of length $L$, the $K$ chosen devices (which will be the leftmost devices in each of the $K$ intervals) will send the data to all the other devices. Due to budget constraints, we cannot send the data to more than $K$ devices initially. We will interpret the points as straight lines in the *time* x *distance* plane. Point $i$ is transformed into the line $y(i,t)=y(i,0)+w(i)\cdot t$, where $y(i,0)=x(i)$ and $w(i)=d(i)\cdot v(i)$. Let's consider a time moment $tc$. At such a moment, the positions of the points are $y(1,tc), \ldots, y(n,tc)$. We consider the points sorted according to their position, i.e. $y(o(1),tc) \leq y(o(2),tc) \leq \ldots \leq y(o(n),tc)$. For the time moment $tc$ it is easy to decide if we can place $K$ intervals of fixed length $L$ which contain all the $n$ points inside them. We place the first interval with its left endpoint at $y(o(1),tc)$. Then we place the next interval with its left endpoint at the position of the leftmost point not contained in any of the previous intervals. We repeat this procedure until every point is part of an interval. If the number $PI$ of intervals we placed is at most $K$, then a solution exists for the time moment $tc$. The following algorithm computes for each position $i$ (in an interval $[p,q]$) the minimum number of intervals $m(i)$ which are required to cover the points $o(i), o(i+1), \ldots, o(q)$. It also computes *next(i)*, the point at which the second interval starts, and *last(i)*, the point where the last interval starts, in an optimal cover of the points *o(i), o(i+1), ..., o(q)*.

**ComputeMinimumNumberOfIntervals(tc, L, p, q):**
*right=q*
**for** *i=q* down to *p* **do**
  **while** *(y(o(right),tc)-y(o(i),tc)>L)* **do** *right=right-1*
  *next(i)=right+1; m(i)=1+(if right=q then 0 else m(right+1))*
  **if** *(right=q)* **then** *last(i)=i* **else** *last(i)=last(right+1)*

The time complexity of the algorithm is linear. Let's consider two sets of time moments: $S_{close} = \{ tc \mid \exists (i,j), i \neq j, s.t. \mid y(i,tc) - y(j,tc) \mid = L$ and $\mid y(i, tc-\varepsilon) - y(j, tc-\varepsilon) \mid \neq L \}$, where $\varepsilon > 0$ is an arbitrarily small constant, and $S_{cross} = \{ tc \mid \exists (i,j), i \neq j, s.t. \ y(i,tc) = y(j,tc) $ and $ y(i, tc-\varepsilon) \neq y(j, tc-\varepsilon) \}$. The earliest time moment *te* when the points can be covered by *K* intervals belongs to the set $S = S_{close} \cup S_{cross} \cup \{0\}$. This can be easily proven in the following way. Let's assume that we run the algorithm described above for *tc=0*, *p=1*, *q=n*. The next time moment *tc'* when the values *m(i)* change (and, thus, the value *m(1)* which is the minimum number of intervals required for covering all the points) is one of the time moments in *S*. The set *S* has cardinality $O(n^2)$. By sorting the time moments in *S* and running the described algorithm for each time moment *tc*, we obtain an $O(n^3)$ solution to our problem; we maintain an array *o* with the order of the points, i.e. $y(o(1),tc) \leq y(o(2),tc) \leq ... \leq y(o(n),tc)$; when we reach a time moment *tc* from the set $S_{cross}$, we need to swap the order of two lines *o(i)* and *o(i+1)* in the array *o* before running the linear algorithm. An interesting question is whether the value *m(1)* can be maintained more efficiently than recomputing it from scratch at every time moment *tc*. An affirmative answer was provided by Dr. M. Patrascu, in a personal communication (July 2008). I will briefly describe his approach here. We will split the lines into *n/k* groups of (approximately) *k* lines each. Each group is composed of lines which are consecutive in the y-ordering (the *o* array). For each group of lines we compute, in the beginning, the values $m_{local}(i)$, $next_{local}(i)$ and $last_{local}(i)$ for each line *o(i)* in the group, having the same meaning as *m(\*)*, *next(\*)* and *last(\*)*, except that they are computed considering only the lines from the group (taking *O(k)* time for each group, using the algorithm presented previously). We also compute in the beginning the values *m(\*)*, *next(\*)* and *last(\*)*, considering all the lines (*[p,q]=[1,n]*). During the algorithm we will not actively maintain the values *m(\*)* and *last(\*)*, but the *next(\*)* values need to be maintained updated. At every time moment *tc* from $S_{cross}$, we just swap the order of the two crossing lines, while the values $m_{local}(*)$, $next_{local}(*)$, $last_{local}(*)$ and *next(\*)* do not change. At every time moment *tc* from $S_{close}$, when two lines *i* and *j* are at distance *L*, we distinguish two cases. Let's assume that *o(a)=i* and *o(b)=j* (*a<b*). We can compute *a* and *b* by maintaining a reverse mapping $o^{-1}$, where $o^{-1}(o(p))=p$ ($1 \leq p \leq n$). In the first case, right before time moment *tc*, the lines *o(a)* and *o(b)* were located at a (vertical) distance smaller than *L*. Thus, in the future, the distance between them will increase. At time *tc*, we need to set *next(a)=b*. All the other *next(p)* values (*p≠a*) remain unchanged. If *o(a)* and *o(b)* are in the same group *G*, then we will recompute the values $m_{local}(i)$, $next_{local}(i)$ and $last_{local}(i)$ for all the lines *o(i)* in group *G*. In the second case, the lines *o(a)* and *o(b)* were located at a (vertical) distance larger than *L* right before time moment *tc*. Thus, right before time *tc*, we have that *next(a)=b*. Since *o(a)* and *o(b)* will be at a distance smaller than *L* in the near future, we need to change the value *next(a)* and set it to *b+1*. Like in the first case, if lines *o(a)* and *o(b)* are located in the same group, we will recompute the $m_{local}(*)$, $next_{local}(*)$ and $last_{local}(*)$ values for all the lines in the group. After updating the computed values, at every time moment *tc*, we need to compute *m(1)*, which is the minimum number of intervals required to cover all the 1D points at the given time moment. We can compute *m(1)* in *O(n/k)* time, as follows. We initialize *m(1)* to $m_{local}(1)$ and a pointer *po* to $last_{local}(1)$. Then, while *po* is not in the last group, we perform the following actions: (1) we set *po* to *next(po)*; (2) we increment *m(1)* by $m_{local}(po)$; (3) we set *po* to $last_{local}(po)$. The time complexity is *O(k+n/k)* for every time moment *tc* from *S*. By choosing $k=O(n^{1/2})$, the overall complexity is $O(n^{2.5})$.

## 6. RELATED WORK AND CONCLUSIONS

In this paper we presented efficient, exact and heuristic algorithms for several offline network optimization problems, like network design, network clustering and QoS improvement. These topics are of great interest in the research community and the discussed problems have several applications in practical settings. Network inverse optimization and network design problems have been studied extensively, due to their large theoretical and practical interest: see, e.g. (Farago et al., 2003) and (Duin et al., 1996). Network improvement problems based on budget constrained network upgrades were presented in (Krumke et al., 1998). Clustering problems similar to those discussed in this paper were addressed in (Chen et al., 2007).